# Coherent, multi-heterodyne spectroscopy using stabilized optical frequency combs

Ian Coddington, William C. Swann, and Nathan R. Newbury
*National Institute for Standards and Technology*

The broadband, coherent nature of narrow-linewidth fiber frequency combs is exploited to measure the full complex spectrum of a molecular gas through multi-heterodyne spectroscopy. We measure the absorption and phase shift experienced by each of 155,000 individual frequency comb lines, spaced by 100 MHz and spanning from 1495 nm to 1620 nm, after passing through a hydrogen cyanide gas. The measured phase spectrum agrees with Kramers-Kronig transformation of the absorption spectrum. This technique can provide a full complex spectrum rapidly, over wide bandwidths, and with hertz-level accuracy.

Optical spectroscopy is a critical tool for our understanding of atomic and molecular structure and has further applications ranging from semiconductor physics to biology. Spectroscopy has remained such a valuable tool in part because it has continued to improve with new methods, such as tunable laser spectroscopy and Fourier transform spectroscopy, and allows for ever larger spectral coverage, finer frequency resolution and accuracy.[1] However, even the best current systems are still limited to ~10 MHz resolution and accuracy and often measure only the absorption without information on optical phase shifts. Stabilized frequency combs offer the potential for a dramatic improvement in broadband spectroscopy by providing both the amplitude and phase across hundreds of nanometers, and with a frequency resolution and accuracy six orders of magnitude better than those of current spectrometers.

Stabilized optical frequency combs are generated from the spectrally broadened output of a mode-locked femtosecond laser and have already revolutionized optical frequency metrology.[2,3] When tightly phase-locked to a narrow optical reference laser, the teeth of a frequency comb are equivalent an enormous number ($>10^5$) of narrowband cw optical oscillators evenly spaced across the spectrum and each with a frequency accuracy equal to that of the reference laser. In practice this accuracy can exceed 1 part in $10^{15}$ or < 1 Hz at infrared frequencies. The obstacle for spectroscopy with stabilized frequency combs is that this impressive accuracy and resolution can be accessed *only* if one overcomes the significant hurdle of separately detecting each individual comb tooth.

Recently two groups have used a combination of spectral filtering and high resolution spectral dispersion to successfully spatially resolve individual modes of a ~GHz frequency comb onto a CCD array. One experiment resolved 2200 comb modes covering a 6.5 THz span,[4] and a second resolved 4000 modes covering a 4 THz span.[5] Frequency combs have also been successfully used in Fourier transform spectroscopy (FTS) and cavity-ring down spectroscopy to enhance sensitivity, but without resolving individual comb lines.[6-9] In this paper we demonstrate an alternate approach of multi-heterodyne spectroscopy[10-13] that allows us to resolve individual comb lines directly by down-converting them into an RF spectrum. This method is more easily scalable to both broader spectra and narrower comb spacing than spatial dispersion techniques. Here, we resolve the phase and amplitude of 155,000 individual comb modes at a spacing of 100 MHz over 15.5 THz, limited only by the source bandwidth.

Multi-heterodyne spectroscopy, pioneered by Keilmann, Van der Weide and coworkers, is an elegant technique in which a second frequency comb serves as a local oscillator (LO).[10-13] This LO comb has a frequency spacing (repetition rate) $f_{r,LO}$ slightly different from that of the signal comb $f_r$, so that the heterodyne beat between the signal and LO comb produces a comb in the rf domain, with separation $\Delta f_r = f_{r,LO} - f_r$. In this rf comb, each tooth corresponds to the heterodyne signal between a specific mode of the signal and LO combs. (See Fig. 1 and Refs. [10-13]). Previous demonstrations of multi-heterodyne spectroscopy in the far-infrared used unstabilized frequency combs and it was not possible to resolve individual comb lines.[10,11,14] As a result, the frequency resolution was limited to tens of gigahertz and no phase information was recorded.

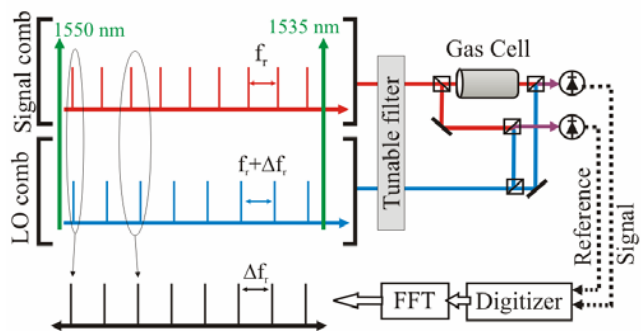

Fig. 1. Simplified schematic of the setup. Signal and LO frequency combs having slightly different tooth spacings are tightly phase-locked to two narrow linewidth laser sources. The signal comb is transmitted through the cell and heterodyned against the LO comb, yielding an RF comb containing information on the amplitude and phase of each tooth in the signal comb. (The laser and rf comb structures are not to scale). Fiber optics transmitted the comb light to two interferometers. A signal interferometer contained a HCN gas while a second reference intereferometer provided a normalization signal.

The challenge in realizing the full potential of multi-heterodyne spectroscopy, is that the relative linewidth between the two combs must, at a minimum, be less than $\Delta f_r$ or the simple rf heterodyne comb of Fig. 1 is swallowed by the overlapping rf noise on each beat. To reach high signal-to-noise ratio (SNR), an even stronger condition is required that the inverse linewidth, i.e. relative coherence time between the two combs, exceed the acquisition time to allow coherent accumulation of the signal. Exploiting recent advances in control of fiber-based modelocked lasers[15,16] we demonstrate multi-heterodyne spectroscopy using two combs with relative linewidths below 1 Hz, limited by our 1 second acquisition time. With two combs tightly phase-locked to common narrow-linewidth cw lasers, we fully resolve the individual rf heterodyne beat from each distinct pair of comb modes. The system yields the full complex (amplitude and phase) spectrum of the sample with a frequency resolution and accuracy set by the cw reference lasers. Furthermore, the broadband SNR increases quadratically in time due to the coherent accumulation of signal, and a spectrum is acquired with only 20 to 2000 pW of power per comb tooth.

It is interesting to compare this technique to standard spectrometry. The resolution of standard grating or FTS spectrometers is limited to the inverse of the differential transit time of the light in the device, yielding values of a few gigahertz and a few megahertz respectively.[1] Similar resolution limits apply to tunable laser spectroscopy due to difficulty in calibrating the instantaneous laser frequency. In contrast, the frequency resolution of multi-heterodyne spectroscopy is limited only by the absolute comb linewidth, which can equal the linewidth of the cw reference laser. For a cw reference laser of 1 Hertz linewidth, the resolution exceeds that of standard grating spectrometers by nine orders of magnitude and that of FTS by six orders of magnitude. The current sample spacing of 100 MHz, set by the laser repetition rate, is similar to that of FTS and is well-matched to the typical few hundred megahertz Doppler-broadened widths of molecular spectra. For Doppler free spectroscopy, the sample spacing can be decreased by sweeping the comb.[4] In addition, as pointed out in Ref. [11], this technique has the potential to acquire broadband spectra more rapidly than can conventional systems, since there are no moving components. For example, $N$ comb modes spanning a bandwidth of $\Delta \nu_{BW} = N f_r$ can be measured in a minimum acquisition time of $t_{acq} = 4 \Delta \nu_{BW} \, f_r^{-2}$, providing $N$ samples with a spacing $f_r$. For $f_r = 100$ MHz, $t_{acq} =$ (0.4 $\Delta \nu_{BW}$) msec/THz.[a] In other words, the full complex spectrum of a sample can be measured over 8 nm of spectral bandwidth in 400 μsec at 100 MHz spacing with a transform-limited frequency accuracy of 2.5 kHz. Rapid acquisition of a complex spectrum opens up interesting possibilities in terms of broadband spectroscopy of dynamical chemical or biological systems.

The basic setup, shown in Fig 1, mirrors that proposed in Ref. [12]. Two frequency combs are generated from mode-locked ring fiber lasers with repetition rates of $f_r =$ 100,016 kHz and $f_{r,LO} =$ 100,017 kHz,[17] the outputs of which are amplified and spectrally broadened in nonlinear fiber. To stabilize the combs we phase-lock two modes of each comb to narrow linewidth cw laser light at 1550 nm and 1535 nm. The 1550 nm light is supplied by a cavity-stabilized cw fiber laser with a hertz-level linewidth and drift of a few hertz per second. Its absolute frequency of ~ 193,361,029,530 kHz is continuously monitored during the experiment through a third self-referenced frequency comb, which is itself monitored against the NIST hydrogen maser ensemble.[15,18] Note that while the reference laser was monitored on the 10 Hz level, drift correction was done only hourly over the data taking period. This leads to a ~10 kilohertz level uncertainty which, while straight forward to remove, was deemed insignificant compared to the gigahertz HCN linewidths. Locked to a tooth near 1535 nm of this self-referenced comb is a 1535 nm cw fiber laser, rendering the 1535 nm laser phase coherent with the 1550 nm laser. (An alternative technique would be to lock both cw lasers to the same optical cavity.) Both cw signals are transmitted over a duplex pair of optical fibers to the multi-heterodyne experiment.

For both the signal and LO frequency combs, a comb tooth is phase-locked to the 1550 nm cw light through feedback to both an intracavity piezoelectric fiber stretcher and an external acousto-optic modulator (AOM) with 100 kHz feedback bandwidth. The AOM transmission limits our source bandwidth to ~125 nm, but a fast intracavity modulator would permit much wider spectral bandwidths. The repetition rate of both combs is stabilized by phase-locking a second comb tooth (~ 19,260 modes away) to the 1535 nm cw light through feedback to the pump power. The four phase-locks have integrated residual phase jitter ranging from 0.27 to 0.47 radians (from 0.1 Hz to 3 MHz), corresponding to a timing jitter of 0.22 to 0.38 fs. The repetition frequency offset, $\Delta f_r \sim 1$ kHz, was controlled through the rf offset of the phase-locks to the 1535 nm light.

The output of each comb is sent through a tunable 3 nm optical grating filter, polarization controller, and, finally, to two interferometers. (See Fig. 1). In the signal interferometer, the signal comb passes through a 15 cm long glass cell filled with 25 Torr of $H^{13}C^{14}N$, and is then combined with the LO comb, generating a beat signal that is digitized at ~100 MHz by a 12 bit ADC. A second reference interferometer without the gas cell provides a normalization signal, detected in parallel. $f_r$, $f_{r,L}$ and $\Delta f_r$ are all counted during the experiment. To generate clean Fourier transform data, the sampling rate is set to an integer multiple of $\Delta f_r$ (100000 in this case). A one second trace corresponds to 128 megasamples, or 1280 heterodyne pulses in the time domain as the two combs overlap.

---

[a] The $N$ rf heterodyne beat notes span $N \Delta f_r$. To avoid aliasing, requires $N \Delta f_r < f_r/2$. Assuming a maximum value of $N \Delta f_r = f_r/4$ to avoid overlapping teeth gives $\Delta \nu_{BW} \Delta f_r = f_r^2/4$. Assuming a minimum data acquisition time $t_{acq} = 1/\Delta f_r$ yields the quoted formula.

Figure 2 shows a rf power spectrum generated from the fast Fourier transform of 1 sec of data after applying a periodic Hanning window of FWHM $(2\Delta f_r)^{-1}$. These data provide the only direct observation of individual frequency-comb modes outside of Ref.s [4,5]. The broadband SNR (peak to noise level) is 35 dB in the center so that any RF noise contamination from adjacent heterodyne beats that could distort the spectrum is negligible. The peak widths are time-bandwith limited to 1 Hz.

Retrieving the optical spectrum from spectra such as that in Figure 2 is simply a question of identifying each discrete rf comb line with the correct optical comb tooth. The signal comb, with repetition rate $f_r$, has modes with optical frequency $\nu_n = \nu_0 + nf_r$, where $\nu_0$ is the calculated frequency of the particular mode locked to the 1550 nm cw reference laser and $n$ is an integer. The LO comb has modes with optical frequency $\nu_m = \nu_0 - \Delta\nu + mf_{r,LO}$, where $\nu_0 - \Delta\nu$ is the calculated frequency of its particular mode locked to the 1550 nm cw reference laser and $m$ is an integer. The heterodyne beat between the $n$th tooth of the signal comb and $m$th tooth of the LO comb occurs at rf frequency $n\Delta f_r - \Delta\nu + (m-n) f_{r,LO}$. (The factor $(n-m)$, when nonzero, is easily calculated from a nanometer-level estimate of the optical filter center.) The phase and squared magnitude of this rf beat corresponds to the relative phase and intensity of the optical comb line at the frequency $\nu_n$.

At each filter setting, the spectrum is normalized by first dividing it by a normalization signal from the reference interferometer. This allows for removal of any laser or filter

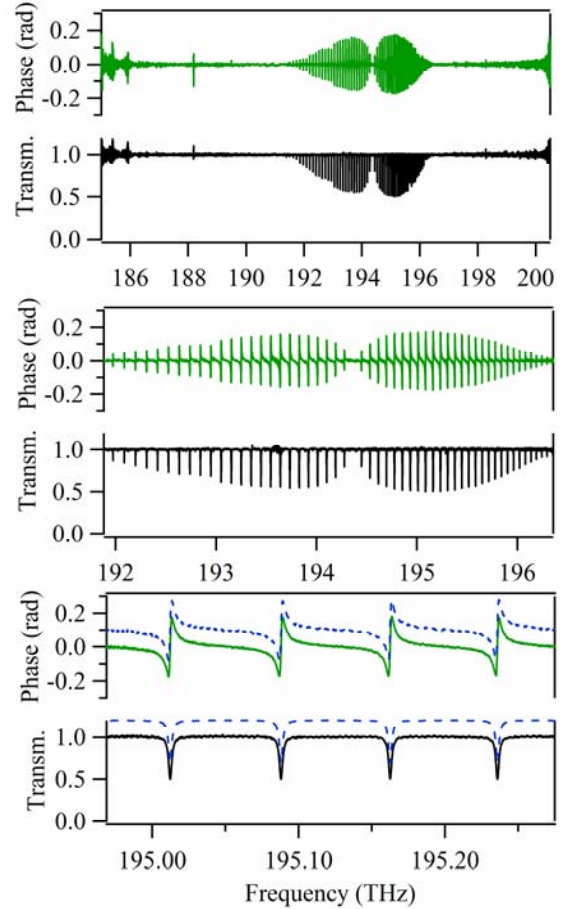

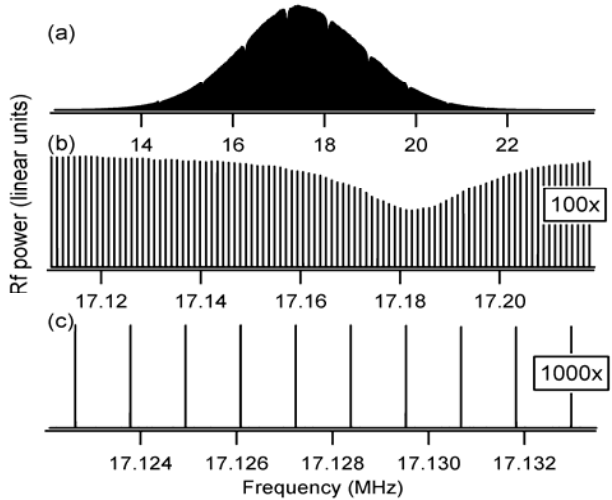

Figure 2: a) Example RF heterodyne power spectrum for 1 second of data at a filter setting of 1536 nm and ~ 200 pW per comb tooth. The -25-dB width of the spectrum spans 10 MHz in the rf, covering 9000 resolved rf heterodyne beat signals between pairs of optical frequency comb lines and 1 THz in the optical domain. The dips are molecular absorption lines. On this scale, the comb modes are too dense to resolve visually. (b) A 100x expanded view near an absorption peak. (c) A 1000x expanded view showing the resolved rf beats, each with a transform-limited 1 Hz FWHM and separated by $\Delta f_r = 1.1$ kHz. These real data compare strikingly well with the schematic of Figure 1.

Figure 3: (a) Measured transmission (black) and phase (green) spectrum for hydrogen cyanide (HCN) spanning 15.5 THz of optical spectrum (from 1495 nm to 1620 nm). There are ~155,000 individual points each corresponding to a single heterodyne beat, such as those shown in Fig. 2. The power per comb tooth varied from 2 nW in the center of the spectrum to 20 pW on the edges of the spectrum. (b) Expanded view covering the 4.5 THz containing the HCN absorption lines. (c) Further expanded view covering 0.3 THz. The measured phase agrees well with the calculated phase (dashed blue line, offset by 0.1 rad) from Kramers-Kronig transformation of the absorption data. The standard deviation of the phase measurement is 1.6 mrad. The measured absorption agrees well

related spectral shape or chirp-related phase shifts between the two combs. Any remaining baseline wander in the amplitude or phase spectra is removed by fitting to a quadratic polynomial. A full spectrum, shown in Fig 3a, is constructed by tuning the optical filter and stitching the resulting spectra together. The filter was scanned in 2 nm increments, and three to five traces, each 1 second long, were averaged. Even at the spectral wings, the individual comb modes are resolved as in Fig 2, although the decreased power per tooth (~20 pW) reduces the signal-to-noise ratio.

The $2\nu_3$ rotational-vibrational band in $H^{13}C^{14}N$ spans 192 to 197 THz and is shown in Figure 3b. Figure 3c gives an expanded view along with previously published data acquired laboriously by use of a slow step-scan of a tunable laser, calibrated against a state-of-the-art wavelength meter, which was itself cross-calibrated against the saturated absorption spectrum of Rb.[19] The SNR of the absorption peaks is ~120, limited by etalon effects from the interferometers and possibly by amplitude noise on the combs. For the ~2300 MHz linewidths and this SNR, the fitted line centers have a ~9 MHz scatter from the published values. Over 47 absorption peaks, the mean line center offset is -1 ± 1.3 MHz from previously published values[19] within the ~1 MHz uncertainty of the published data. Figure 3c shows the measured phase spectrum across the same spectral region, along with the calculated phase from the Kramers-Kronig transformation of the absorption data.[1] Although the absorption lines are widely spaced, the phase delay caused by different absorption peaks overlap with each other. The 2 mrad scatter on the phase measurements corresponds to a 1.6 attosec timing jitter between comb lines. This level of phase stability is promising for a variety of time domain applications such optical waveform characterization, distance metrology or direct dispersion measurements.

In conclusion, we have shown that multi-heterodyne spectroscopy with stabilized frequency combs can be used to measure the full absorption and phase spectrum of a sample over 15.5 THz with 155,000 individual measurements. Even broader spectra would be possible with octave-spanning combs. Furthermore higher sensitivities could be reached by including a White cell or buildup cavity to increase the effective path length through the cell.[5, 6] Other applications of this technique include rapid spectroscopy of dynamical systems, broadband measurements of the transmission and dispersion of optical systems, and coherent time-domain pump-probe experiments. This last application would exploit the rare capability of multi-heterodyne spectroscopy to provide the full amplitude and phase across a wide spectrum. A sample's time domain response to the incoming pulse train could be characterized with high timing resolution and high signal to noise ratio through coherent signal averaging.

We acknowledge assistance in the data analysis from Ljerka Nenadovic, and helpful discussions with Mike Ziemkiewicz, Danielle Braje and Scott Diddams.